# In-Building Wideband Partition Loss Measurements at 2.5 GHz and 60 GHz

Christopher R. Anderson, *Student Member*, and Theodore S. Rappaport, *Fellow, IEEE*

*Abstract*— **This paper contains measured data and empirical models for 2.5 & 60 GHz in-building propagation path loss and multipath delay spread. Path loss measurements were recorded using a broadband sliding correlator channel sounder which recorded over 39,000 Power Delay Profiles (PDPs) in 22 separate locations in a modern office building. Transmitters and receivers were separated by distances ranging from 3.5 to 27.4 meters, and were separated by a variety of obstructions, in order to create realistic environments for future single-cell-per-room wireless networks. Path loss data is coupled with site-specific information to provide insight into channel characteristics. These measurements and models may aid in the development of future in-building wireless networks in the unlicensed 2.4 GHz and 60 GHz bands.**

## I. INTRODUCTION

Over the past decade, the market for wireless service has grown at an unprecedented rate. The industry has expanded from cellular phones and pagers to Personal Communication Systems (PCS), wireless local area networks (WLANs), and broadband wireless services that can provide voice, data, and full-motion video in real time [1]. In order for the visions of 3rd and 4th generation of wireless communication standards to be realized, system design engineers must have a thorough understanding of the wireless channels in which these devices operate.

In recent years, there has been an increasing interest in providing broadband communications in the 2.4 GHz ISM band and the 60 GHz unlicensed band for WLANs. In particular, the propagation characteristics of the 60 GHz band provides the promise of high spatial frequency reuse, with low-power transmitters operating in a single-cell-per-room

Manuscript received May 11, 2002, revised December 2, 2002. This work was supported in part by National Science Foundation Grant Number EIA-9974960.

C. R. Anderson is with the Mobile and Portable Radio Research Group, Virginia Polytechnic Institute and State University, Bradley Department of Electrical & Computer Engineering, 432 Durham Hall, Blacksburg, VA 24061-0350 (phone: 540-231-2927; fax: 540-231-2968; e-mail: chris.anderson@vt.edu).

T. S. Rappaport., is with the Wireless Networking and Communications Group, The University of Texas at Austin, Department of Electrical & Computer Engineering, 433A ENS Building, Austin, TX 78712-1084 (e-mail: wireless@mail.utexas.edu).

configuration [2], [3], called "femtocellular" [2]. Such a system will provide high data-rate services for densely populated buildings, carrying many times more traffic than current wireless networks. While spectrum in the 2.4 and 60 GHz bands has been available for several years, there is a lack of comparisons between the indoor propagation characteristics in these two bands—and it is unclear as to how the penetration losses vary for various objects encountered in a modern office building. Numerous propagation studies have been performed at cellular (900 MHz), PCS (1900 MHz), U-NII (5-6 GHz), LMDS (28-31 GHz), and millimeter-wavelength (60 GHz) frequencies (for example, [4]—[12]) however, little is know about the differences between the 2.4 and 60 GHz bands.

Davies et. al., conducted one of the earliest studies of the differences between microwave and millimeter-wavelength frequencies [11], which investigated wideband propagation effects encountered in a single-cell-per-room environment. Davies et. al. observed that RMS delay spreads for 60 GHz propagation was significantly lower than for 1.7 GHz propagation, and was attributable primarily to three important propagation phenomenon: (1) variation of the electrical parameters (reflection coefficient, conductivity, etc.) of building materials with frequency, (2) significant attenuation of 60 GHz signals by building materials, and (3) the use of omnidirectional antennas at 1.7 GHz and more directional horn antennas at 60 GHz [11]. Another study on the spatial variation of received power in a single building at 900 MHz and 60 GHz was conducted by Alexander and Pugliese [13]. Alexander and Pugliese observed that a 900 MHz signal was capable of covering multiple rooms in a building, whereas 60 GHz signals were generally confined to a single room, due to significantly higher attenuation of 60 GHz signals by building materials [13].

Most of the available literature has so far concentrated on investigating penetration loss into buildings, rather than from obstructions inside buildings. Several propagation studies show that penetration loss of various building materials increases as the transmission frequency increases. Zhang and Hwang as well as Golding and Livine show how penetration loss in various building materials increases over the frequency range of 900 MHz—18 GHz and 20—50 GHz, respectively [14], [15]. Additionally, penetration losses for building materials at various frequencies between 5 and 60 GHz are reported in [4], [14], [16]—[21], and a general increase in penetration loss as frequency increases can be observed.

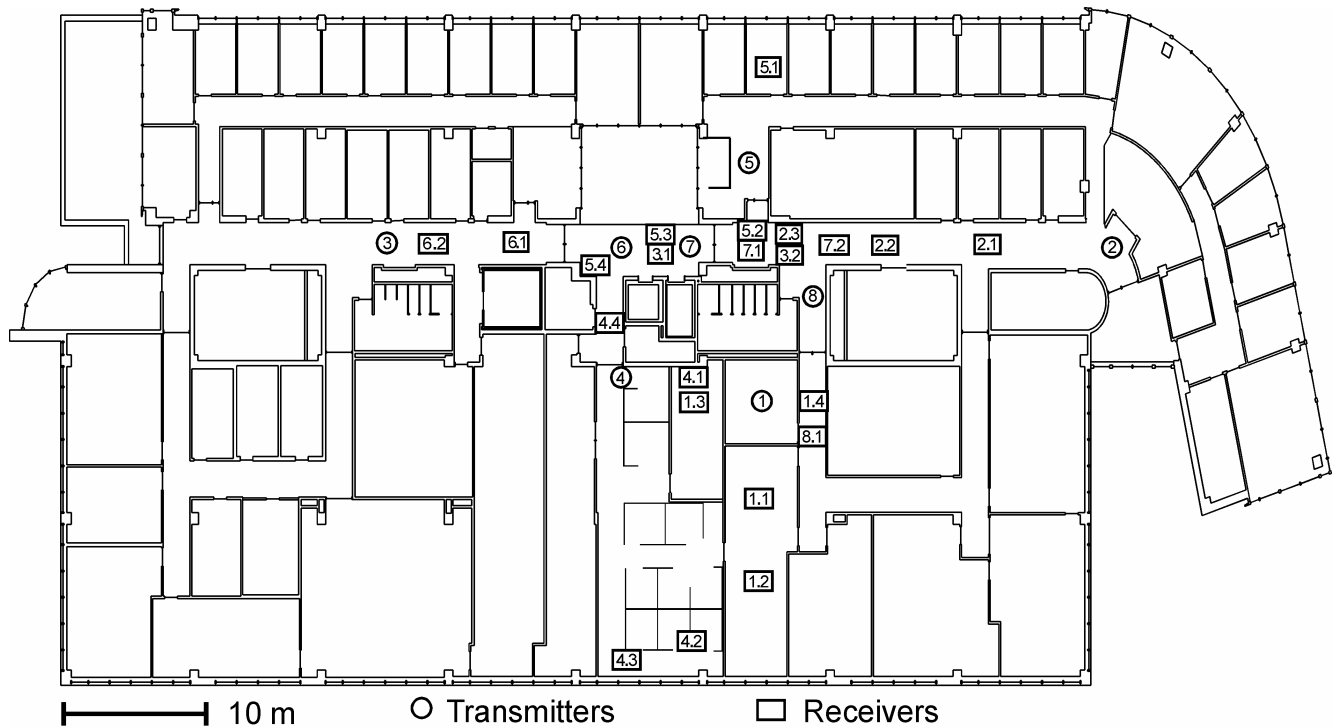

**Fig. 1.  Map of the 4th floor of Durham Hall at Virginia Tech, with transmitter and receiver locations identified**

These data compare favorably to the penetration losses reported in Section III of this paper.

Using measured penetration losses, Durgin, et. al., Nobles, et. al., and Karlsson, et. al., have developed indoor propagation models at 5.8 GHz, 17 & 60 GHz, and 5.0 GHz, respectively, to predict path loss based on the number and types of obstructions encountered between transmitter and receiver [4], [22], [23]. These models can also be used to characterize the site-specific nature of emerging femtocellular systems, and may be used in ray-tracing algorithms to predict network coverage and throughput.

This paper presents results on signal propagation in a modern four story office building built in 1998. The measurement campaign involved recording wideband power delay profiles (PDPs) in a typical indoor office environment. Measurements at both 2.5 and 60 GHz were recorded in 22 separate locations on the 4th floor inside the building, requiring over 39,000 individual PDPs. From measurements, we develop a large-scale path loss model based on log-normal propagation, as well as a more site-specific model based on the Partition-Dependant Propagation Model described in [4].

Section II discusses the experimental hardware, setup, methodology, and measurement campaign. Sections III—IV summarizes the results, presents models for in-building path loss and propagation loss, and presents conclusions.

## II. EXPERIMENTAL SETUP

The following section describes the methodology for measuring path loss and penetration loss. Definitions of path loss and penetration loss as well as descriptions of measurement procedures and sites are included.

### A. Description of Measurement Procedure and Locations

Eight separate transmitter and 22 separate receiver locations were selected on the 4th floor of Durham Hall on the Virginia Tech campus. The measurement sites were chosen to be representative of a broad range of typical femtocellular propagation environments in a work setting, where a low power transmitter will serve a single room or portion of a floor. Durham Hall was completed in 1998, with a foundation and framework made from steel reinforced concrete, with interior sheetrock and concrete cinderblock walls, ceramic tile and carpeted floors, and suspended panel ceilings. Fig. 1 illustrates the building floor plan and identifies transmitter and receiver locations for this measurement campaign. Measurements were grouped into eight different segments, based on transmitter location, and numbered based on both transmitter and receiver location. These measurements look specifically at the wideband propagation effects that may be encountered in a typical office building, with transmitter and receiver locations chosen to provide line-of-sight (LOS), non line-of-sight (NLOS), and cluttered propagation environments.

A broadband vector sliding correlator channel sounder, developed in [24] was used to record wideband PDPs at all measurement locations. A basic block diagram of the channel sounder is shown in Fig. 2. The channel sounder utilized an 11-bit pseudo-random noise code running at 400 MHz, with GPS disciplined oscillators generating a highly stable frequency reference at transmitter and receiver, providing the channel sounder with a multipath temporal resolution of 2.5 nanoseconds. Two different RF front ends were utilized, one for the 60 GHz measurements and the other for the 2.5 GHz measurements.

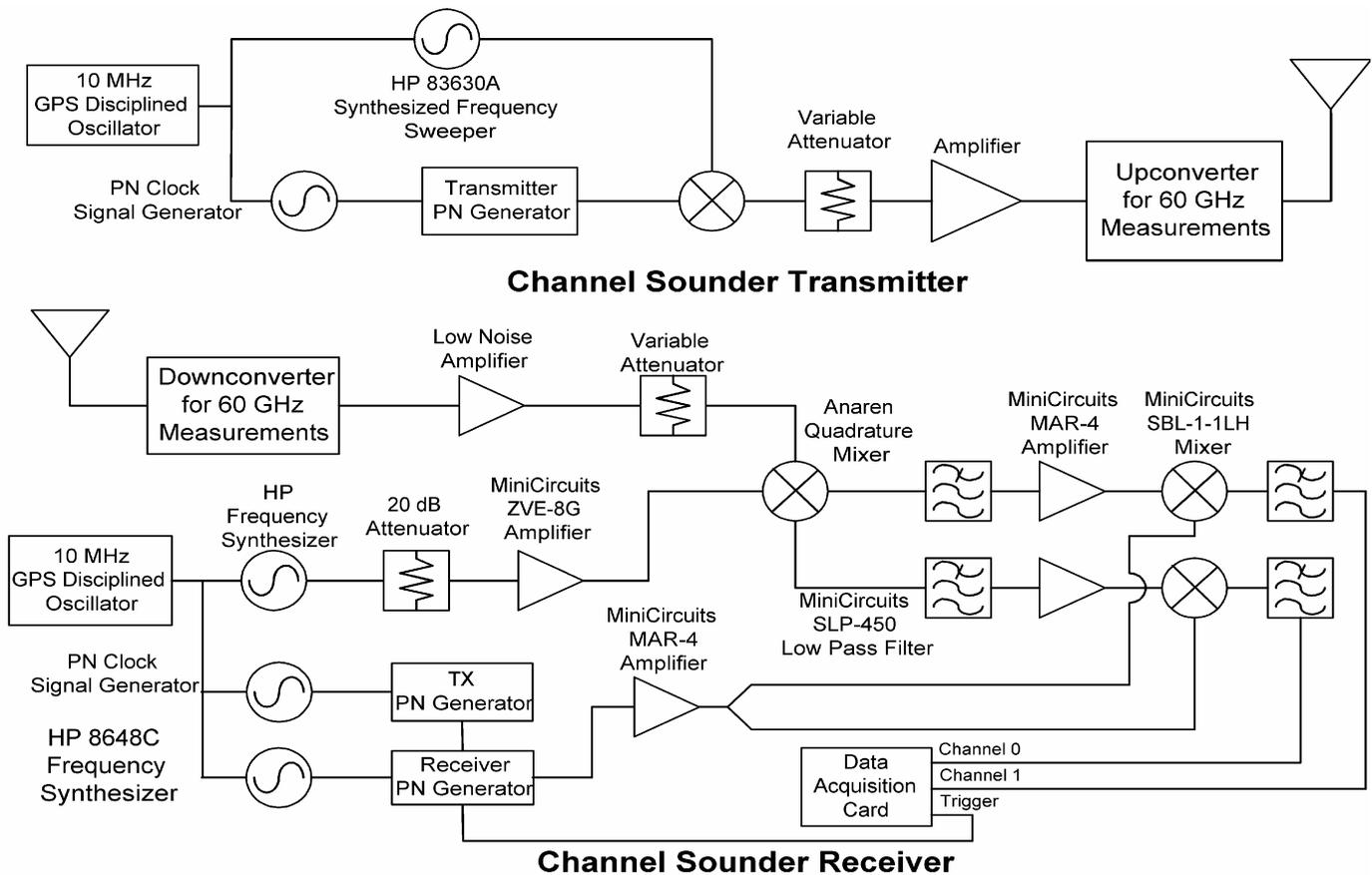

**Fig. 2. Block diagram of the broadband sliding correlator channel sounder used in the measurement campaign**

For the 60 GHz measurements, the transmitter and receiver utilized pyramidal horn antennas which had a gain of 25 dBi and first-null beamwidths of 50°. These high-gain horn antennas were used in order to overcome the considerable amount of path loss at 60 GHz. Transmitter output power (as measured at the input to the antenna) was set at -10 dBm. The relatively low transmitter power was necessary in order to maintain the linear operation of the transmitter power amplifier, as well as avoid saturating the receiver low noise amplifier at short link distances. Additionally, the low output power (+15 dBm EIRP) emulated femtocellular systems where it is desired to contain a transmitted signal within the boundaries of a single room. For the 2.5 GHz measurements, transmitter and receiver utilized omnidirectional biconical antennas with a 6 dBi gain. These lower gain omnidirectional antennas were used due to their compact physical size, as well as to emulate 2.4 GHz WLANs operating with omnidirectional antennas. Transmitter output power (as measured at the input to the antenna) was set at 0 dBm.

For both configurations, transmitter and receiver antennas were vertically polarized, and heights were nominally set at 1.2 meters, with the exception of transmitter location 4 where the antenna height was increased to 2.4 meters. A laptop computer was used to record inphase and quadrature components of the complex impulse response, and software post-processing generated the PDPs.

Calibration was performed at the beginning and end of both 2.5 and 60 GHz measurements to ensure the accuracy of the measurement system. Overall system gain and reference path loss were calculated from these calibration runs, and consistently provided an accuracy of ±1.3 dB and a repeatability of ±1.5 dB. The channel sounder had a measurement dynamic range of 30 dB with a maximum and minimum received power level of -20 dBm and -85 dBm, respectively.

In this measurement campaign, a series of power delay profiles were recorded along a linear measurement track. The track has a positioning accuracy of ±10 μm [5], which is critically important since the wavelength at 60 GHz is only 5 millimeters. For both 2.5 GHz and 60 GHz, two types of track measurements were performed, a parallel track measurement and a perpendicular track measurement. **Parallel Track Measurements** consisted of orienting the track parallel to the line connecting transmitter to receiver. **Perpendicular Track Measurements** consisted of orienting the track perpendicular (transverse) to the line connecting transmitter to receiver. For both parallel and perpendicular measurements, power delay profiles were recorded at 40 positions along the track with a separation between successive measurements of $\frac{\lambda}{4}$, so that the receiver traveled a total distance of $10\lambda$.

Additionally, at 60 GHz a second set of parallel and perpendicular track measurements were performed where power delay profiles were recorded at 10 positions along the track with a separation between successive measurements of $10\lambda$, so that the receiver traveled a total distance of ½ meter ($100\lambda$).

All power delay profiles at a given measurement location were then linearly averaged together to create a composite *local area power delay profile*. All measured local area power delay profiles may be found in [24].

### B. Definition of Path Loss

To measure path loss, the channel sounder records wideband power delay profiles at 2.5 GHz and 60 GHz. Narrowband received power fluctuates over a small area due to multipath-induced fading, however, averaging power over a local area yields a reliable estimate for the local average received power independent of signal bandwidth [16]. Additionally, narrowband power can be calculated from a wideband PDP using the following relationship [25]

$$P_R\,(dBm) = P_{cal}\,(dBm) + 10\log_{10}\left(\frac{\sum_r G_r\Delta\tau_r}{\sum_r G_{cal}\Delta\tau_{cal}}\right) \quad (1)$$

where $\sum_r G_r\Delta\tau_r$ is the integrated power in a given PDP (area under the PDP). The term $\sum_r G_{cal}\Delta\tau_{cal}$ is the integrated power in a PDP from a calibration run and is assigned the known input power $P_{cal}$.

For the purpose of this paper, path loss (PL) is defined as the *ratio of the effective transmitted power to the received power, calibrating out all system losses, amplifier gains, and antenna gains*. All measured path loss values reported in this paper have the antenna gains subtracted out—providing path loss that would be experienced if isotropic antennas were used on the transmitter and receiver, even though directional antennas were actually employed on the measurement system. The measured path loss from transmitter to receiver is then given by

$$PL = P_T + G_T + G_R - P_{REC} \quad (2)$$

where $PL$ is the measured path loss in dB, $P_T$ and $P_{REC}$ are the transmitter and receiver powers in dBm, and $G_T$ and $G_R$ are the transmitter and receiver antenna gains in dBi (dB of gain with respect to an ideal isotropic antenna [26]).

Free space path loss between transmitter and receiver, assuming isotropic antennas, is given by:

$$PL_{FS} = -10\log_{10}\left(\frac{\lambda^2}{(4\pi d)^2}\right) \quad (3)$$

Where $PL_{FS}$ is the free-space path loss in dB, $\lambda$ is the wavelength (12 cm at 2.5 GHz, 0.5 cm at 60 GHz), and $d$ is the transmitter-receiver separation distance in meters.

Additionally, it has been well documented in the literature that large-scale path loss for an arbitrary transmitter-receiver separation distance is distributed log-normally about the distance-dependant mean, with path loss given by [16]

$$PL(d) = PL(d_0) + 10n\log_{10}\left(\frac{d}{d_0}\right) + X_\sigma \quad (4)$$

Where $PL(d)$ is the path loss in dB at a particular distance $d$ in meters from the transmitter, $PL(d_0)$ is the path loss in dB at a close-in free space reference distance $d_0$, $n$ is the path loss exponent, and $X_\sigma$ is a zero-mean Gaussian random variable with standard deviation of $\sigma$ dB.

### III. OBSERVATIONS, TRENDS, AND SITE-SPECIFIC MEASUREMENT RESULTS

This section presents path loss data recorded from all measurement sites described in Section II. Path loss values are reported and compared to theoretical free-space path loss as predicted by (3).

### A. Summary of Results

Tables I and II summarize the results of the 39,600 power delay profiles recorded during this measurement campaign. The tables are ordered by receiver location number. Listed in the tables are the *Theoretical Free Space Path Loss*, *Local Area Average Path Loss, Local Area Min/Max Path Loss*, and *Local Area Min/Max/Average RMS Delay Spread.*

*Free Space Path Loss* is the path loss from transmitter to receiver calculated using (3). Local Area Average Path Loss is the measured path loss averaged over all receiver positions (for both parallel and perpendicular track measurements) at a given measurement location, and is a spatial linear average value for path loss at that measurement location. Because the received power measured by the channel sounder includes the effects of the transmitter and receiver antennas, the antenna gains were subtracted out from the measurements, in order to provide a meaningful comparison with the theoretical free-space path loss.

*Local Area Min/Max Path Loss* is the minimum and maximum path loss among all PDPs recorded at a particular receiver location. Similarly, the *Local Area Min/Max/Avg $\sigma_\tau$* is the minimum, maximum, and average RMS delay spread among all PDPs recorded at a particular measurement location.

**TABLE I**
SUMMARY OF 2.5 GHz MEASUREMENT RESULTS ON A
SINGLE FLOOR OF A MODERN OFFICE BUILDING

| Location | Link Distance (m) | Free Space PL (dB) | Local Area Avg. PL (dB) | Local Area Min / Max PL (dB) | Local Area Min / Max / Avg. $\sigma_\tau$ (ns) |
|---|---|---|---|---|---|
| 1.1 | 5.4 | 55 | 64 | 63 / 64 | 32 / 42 / 37 |
| 1.2 | 9.2 | 60 | 64 | 64 / 65 | 23 / 29 / 26 |
| 1.3 | 4.7 | 54 | 55 | 54 / 56 | 19 / 22 / 20 |
| 1.4 | 3.5 | 51 | 54 | 53 / 54 | 16 / 24 / 20 |
| 2.1 | 7.8 | 58 | 62 | 60 / 64 | 7 / 19 / 13 |
| 2.2 | 16.2 | 65 | 63 | 61 / 65 | 16 / 23 / 22 |
| 2.3 | 22.9 | 68 | 67 | 67 / 68 | 8 / 11 / 10 |
| 3.1 | 18.2 | 66 | 63 | 62 / 64 | 37 / 48 / 43 |
| 3.2 | 27.4 | 69 | 66 | 64 / 68 | 26 / 36 / 31 |
| 4.1 | 6.0 | 56 | 64 | 64 / 65 | 17 / 33 / 25 |
| 4.2 | 13.0 | 63 | 65 | 64 / 65 | 22 / 32 / 27 |
| 4.3 | 13.6 | 63 | 64 | 63 / 65 | 21 / 22 / 21 |
| 4.4 | 4.7 | 54 | 64 | 64 / 65 | 29 / 31 / 30 |
| 5.1 | 4.5 | 54 | 54 | 53 / 54 | 19 / 20 / 20 |
| 5.2 | 12.2 | 62 | 73 | 72 / 73 | 18 / 24 / 21 |
| 5.3 | 7.7 | 58 | 72 | 71 / 72 | 27 / 28 / 27 |
| 5.4 | 3.9 | 52 | 55 | 55 / 56 | 17 / 18 / 18 |
| 6.1 | 7.6 | 58 | 68 | 67 / 68 | 15 / 16 / 16 |
| 6.2 | 17.1 | 65 | 68 | 67 / 68 | 24 / 30 / 26 |
| 7.1 | 5.5 | 55 | 67 | 66 / 68 | 15 / 16 / 16 |
| 7.2 | 10.4 | 61 | 68 | 67 / 68 | 9 / 10 / 9 |
| 8.1 | 5.5 | 55 | 64 | 63 / 64 | 21 / 22 / 21 |

**TABLE II**
SUMMARY OF 60 GHz MEASUREMENT RESULTS ON A
SINGLE FLOOR OF A MODERN OFFICE BUILDING

| Location | Link Distance (m) | Free Space PL (dB) | Local Area Avg. PL (dB) | Local Area Min / Max PL (dB) | Local Area Min / Max / Avg. $\sigma_\tau$ (ns) |
|---|---|---|---|---|---|
| 1.1 | 5.4 | 83 | 98 | 97 / 99 | 15 / 20 / 18 |
| 1.2 | 9.2 | 87 | 103 | 101 / 105 | 18 / 22 / 20 |
| 1.3 | 4.7 | 81 | 93 | 91 / 94 | 16 / 18 / 17 |
| 1.4 | 3.5 | 79 | 82 | 81 / 83 | 1 / 4 / 3 |
| 2.1 | 7.8 | 86 | 73 | 72 / 74 | 3 / 8 / 6 |
| 2.2 | 16.2 | 92 | 78 | 76 / 87 | 7 / 9 / 8 |
| 2.3 | 22.9 | 95 | 98 | 95 / 104 | 6 / 9 / 8 |
| 3.1 | 18.2 | 93 | 89 | 88 / 90 | 11 / 17 / 14 |
| 3.2 | 27.4 | 97 | 99 | 97 / 100 | 8 / 12 / 9 |
| 4.1 | 6.0 | 84 | 94 | 89 / 98 | 10 / 16 / 11 |
| 4.2 | 13.0 | 90 | 99 | 97 / 101 | 9 / 18 / 14 |
| 4.3 | 13.6 | 91 | 91 | 89 / 97 | 10 / 19 / 15 |
| 4.4 | 4.7 | 81 | 89 | 88 / 90 | 6 / 22 / 13 |
| 5.1 | 4.5 | 81 | 81 | 76 / 83 | 6 / 12 / 8 |
| 5.2 | 12.2 | 90 | 96 | 94 / 97 | 3 / 12 / 8 |
| 5.3 | 7.7 | 86 | 87 | 85 / 87 | 5 / 10 / 8 |
| 5.4 | 3.9 | 80 | 80 | 73 / 83 | 3 / 6 / 4 |
| 6.1 | 7.6 | 86 | 89 | 88 / 91 | 2 / 12 / 6 |
| 6.2 | 17.1 | 93 | 103 | 97 / 107 | 19 / 30 / 25 |
| 7.1 | 5.5 | 83 | 94 | 93 / 95 | 9 / 14 / 12 |
| 7.2 | 10.4 | 88 | 99 | 97 / 99 | 6 / 12 / 9 |
| 8.1 | 5.5 | 83 | 85 | 84 / 86 | 17 / 24 / 19 |

Note that very little difference exists between the maximum and minimum path loss and the maximum and minimum RMS delay spreads, due to minimal fading of multipath components over the local area, a result also demonstrated theoretically in [16].

Figs. 3a and 3b are scatter plots of all measured path loss values versus distance for the 2.5 GHz and 60 GHz measurements. A Minimum Mean Square Error analysis [16] was applied to the measured data to determine the path loss exponent as expressed in (4). The resulting path loss exponent for 2.5 GHz was $n = 2.4$, with a standard deviation of $\sigma = 5.8\,\text{dB}$ and for 60 GHz the path loss exponent was $n = 2.1$ with $\sigma = 7.9\,\text{dB}$, which are within the ranges for in-building same-floor propagation reported in [6], [16], [22], [25], [27].

### B. Partition Based Path Loss Analysis and Channel Model

In propagation analysis the path loss exponent, $n$, is useful for predicting large-scale propagation effects. However, the path loss exponent model is inadequate at predicting site-specific propagation effects, such as reflection, diffraction, or penetration losses caused by a particular building layout, construction materials, furniture, etc. A more refined model uses *partition-dependant attenuation factors* [4], [28], which assumes free space propagation ($n=2$) with additional path loss incurred based on the number and type of objects (such as walls or doors) intersected by a single ray drawn from transmitter to receiver. A pseudo deterministic method for

determining the received power in such an environment is given by [4]

$$P_R\left(d\right) = P_T + G_T + G_R - 20\log_{10}\left(\frac{4\pi d}{\lambda}\right) - \sum_{i=1}^{N} a_i X_i \quad (5)$$

where $P_R\left(d\right)$ is the received power in dBm at a particular distance $d$ in meters from the transmitter, $P_T$ is the transmitter power in dBm, $X_i$ are the attenuation values in dB for the $i$th partitions intersected by a line drawn from the transmitter to the receiver, and $a_i$ are the number of times the ray intersects each type of partition (i.e. $a_1$ intersections with partition $X_1$, $a_2$ intersections with partition $X_2$, and so forth). Note that in the case of free space propagation with no partitions between transmitter and receiver, path loss calculated from (5) will be identical to path loss calculated from (3). Also, the partition based channel model works well for short transmitter-receiver separations, provided there are a small number of multipath scatterers in the environment. If a significant amount of

PARTITION LOSSES (LOSS IN EXCESS OF FREE SPACE) AT 2.5 & 60 GHz
ON THE 4TH FLOOR OF DURHAM HALL, VIRGINIA TECH

|  |  | Drywall | Office Whiteboard | Clear Glass | Mesh Glass | Clutter |
|---|---|---|---|---|---|---|
| Number of Measurements at Each Frequency | | 7 | 4 | 4 | 4 | 4 |
| Material Thickness (cm) | | 2.5 | 1.9 | 0.3 | 0.3 | -- |
| **2.5 GHz** | Average Measured Attenuation (dB) | 5.4 | 0.5 | 6.4 | 7.7 | 2.5 |
| | Measurement Standard Deviation (dB) | 2.1 | 2.3 | 1.9 | 1.4 | 2.2 |
| | Normalized Average Attenuation (dB/cm) | 2.1 | 0.3 | 20.0 | 24.1 | -- |
| **60 GHz** | Average Measured Attenuation (dB) | 6.0 | 9.6 | 3.6 | 10.2 | 1.2 |
| | Measurement Standard Deviation (dB) | 3.4 | 1.3 | 2.2 | 2.1 | 1.8 |
| | Normalized Average Attenuation (dB/cm) | 2.4 | 5.0 | 11.3 | 31.9 | -- |

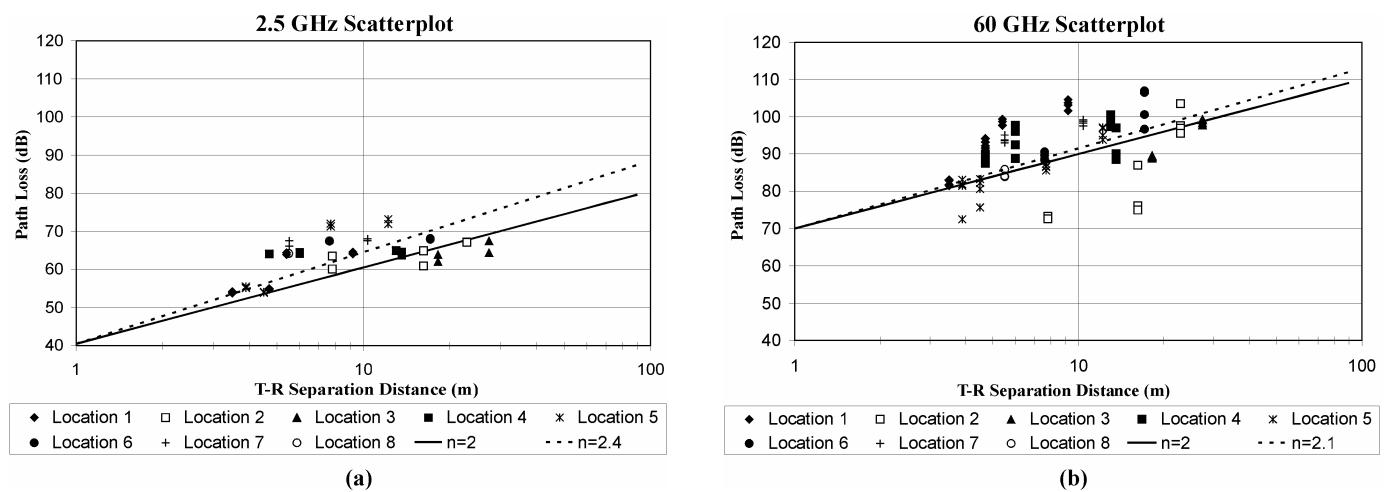

Fig. 4.  Scatterplot of all Measured Path Loss Values (Excluding Antenna Gains) on a Single Floor of a Modern Office Building for  (a) 2.5 GHz with Transmitter and Receiver using Omnidirectional Biconical Antennas with 6dBi Gain, and (b) 60 GHz with Transmitter and  Receiver using Pyramidal Horn Antennas with 25dBi Gain and  50° First-Null Beamwidth

the received power comes from multipath, then the partition based model loses its physical significance.

By looking at the building floor plan (Fig. 1), partitions that existed between each transmitter and receiver link were placed in five separate categories:

1. Drywall:  2 sheets of standard ½ inch thick sheetrock wallboard.

2. Office Whiteboard:  Standard office dry-erase melamine whiteboard, attached to ½ inch thick plywood backing.

3. *Clear Glass*:  ⅛ inch thick clear glass which is untextured and unreinforced.

4. *Mesh Glass*:  ⅛ inch thick clear glass which has been reinforced with interlacing 24 gauge wires configured in a rectangular grid with openings of ½ inch × ½ inch.

5. *Clutter*:  Objects that encroached into the first Fresnel zone but did not directly block the LOS signal from transmitter to receiver.  Clutter includes office furniture such as chairs, desks, bookcases, and filing cabinets, in addition to soft partitions that did not extend to the ceiling.

A summary of all partition attenuation factors at 2.5 GHz and 60 GHz is shown in Table III, with the attenuation values representing loss in excess of free space, i.e., loss induced by the partition in addition to the ideal free space path loss ($n = 2$), and represent a composite average calculated from all measurements containing a particular partition.  Additionally, to ensure that partition attenuation values could be compared in a meaningful way, all attenuation values were normalized to dB per centimeter of material thickness.  The measurement standard deviation represents the variance of the measured partition losses for all transmitter-receiver links containing a particular partition.

### C. Discussion of Results

Comparing Figs 3a and 3b, it can be seen that when using highly directional antennas at 60 GHz, the large-scale mean path loss is very close to free-space, but the spread of values about the mean is higher than at 2.5 GHz. It must also be noted that the RMS Delay Spread at 2.5 GHz is significantly higher than at 60 GHz, which was caused by the use of omnidirectional biconical antennas at 2.5 GHz and horn antennas at 60 GHz.

Comparing the measured partition losses for 2.5 GHz and 60 GHz in Table III, the attenuation of drywall, whiteboard, and mesh glass increases from 2.5 GHz to 60 GHz, however the attenuation of clear glass and clutter decreases. The reason for the decrease in attenuation from clutter is intuitive, since the first Fresnel zone at 60 GHz is significantly smaller than at 2.5 GHz, and therefore fewer objects in the environment are capable of perturbing the LOS signal. It is not known why the attenuation of Clear Glass, as reported in Table III, decreases from 2.5 GHz to 60 GHz.

## IV. CONCLUSIONS

This paper presented the results of a measurement campaign and detailed analysis of in-building 2.5 GHz and 60 GHz wireless channels. Measurements were analyzed in context with site-specific information, and results include local average path loss and partition loss values for a variety of materials encountered in an office or laboratory building. The measured path loss exponents and partition losses agree well with results published in the literature. The data presented in this paper highlights several differences between propagation at 2.5 & 60 GHz:

1. At 60 GHz, propagation is more ray-like [13], and the structure and composition of partitions in the environment (such as the use of metal studs in interior walls) can have a significant impact on multipath delay spreads, whereas 2.5 GHz signals are less sensitive to such details.

2. High levels of attenuation for certain building materials, in addition to significantly higher free-space path loss may aid in keeping 60 GHz signals confined to a single room. As a result, 2.5 GHz systems would be more effective at covering several rooms or a portion of a building floor, whereas 60 GHz systems would be ideal for a femtocellular network consisting of a single-cell-per-room approach.

3. The higher partition losses at 60 GHz effectively restrict received multipath components to reflectors within a single room [11]. As a result, RMS delay spreads are very low and leads to the possibility of providing very high data rate communications within a single room.

The partition based path loss model, developed in [4], was applied to measured path loss data at 2.5 & 60 GHz. The partition based model provides fast results for predicting path loss with a minimal amount of calculation, and is suitable for incorporation into software site modeling and planning tools, which will aid in the design, site planning, and deployment of future in-building wireless femtocellular networks.

## REFERENCES

[1] T. S. Rappaport, A. Annamalai, R. M. Buehrer, and W. H. Tranter, "Wireless communications: past events and future perspectives," *IEEE Communications Magazine 50th Anniversary Issue*, pp. 148-161, May 2002.W.-K. Chen, *Linear Networks and Systems* (Book style). Belmont, CA: Wadsworth, 1993, pp. 123–135.

[2] G. Vannucci and R. S. Roman, "Measurement results on indoor radio frequency reuse at 900 MHz and 18 GHz," *IEEE 3rd International Conference on Personal, Indoor, and Mobile Radio Communications*, pp. 308-314, October 1992.

[3] D. Molkdar, "Review on radio propagation into and within buildings," *IEE Proceedings—Microwaves, Antennas, and Propagation*, Vol. 138, No. 1, pp. 61-73, February 1991.

[4] G. Durgin, T. S. Rappaport, and H. Xu, "Measurements and models for radio path loss and penetration loss in and around homes and trees at 5.85 GHz," *IEEE Transactions on Communications*, vol. 46, No. 11, November 1998.

[5] G. D. Durgin, "Theory of stochastic local area channel modeling for wireless communications," Ph. D. Dissertation, Virginia Polytechnic Institute and State University, http://scholar.lib.vt.edu/theses/index.html, December 2000.

[6] H. Xu, V. Kukshya, and T. S. Rappaport, "Spatial and temporal characteristics of 60-GHz indoor channels," *IEEE Journal on Selected Areas in Communications*, vol. 20, No. 3, pp. 620-630, April 2002.

[7] W. J. Tanis, II, and G. J. Pilato, "Building penetration characteristics of 880 MHz and 1922 MHz radio waves," *Proceedings IEEE 43rd Vehicular Technology Conference*, Secanus, NJ, May 1993, pp. 2006-209.

[8] A. F. de Toledo, and A. M. D. Turkmani, "Propagation into and within buildings at 900, 1800, and 2300 MHz," *Proceedings IEEE 42nd Vehicular Technology Conference*, Denver, CO, May 1992, vol. 2, pp. 633-655.

[9] E. H. Walker, "Penetration of radio signals into buildings in the cellular radio environment," *Bell System Technical Journal*, vol. 62, no. 9, pp. 2719-2734, Nov. 1983.

[10] Y. P. Zhang, and Y. Hwang, "Time delay characteristics of 2.4 GHz band radio propagation channels in room environments," *Proceedings IEEE 5th Personal, Indoor and Mobile Radio Communications*, The Hague, Netherlands, September 1994, vol. 1, pp. 28-32.

[11] R. Davies, M. Bensebti, M. A. Beach, and J. P. McGeehan, "Wireless propagation measurements in indoor multipath environments at 1.7 GHz and 60 GHz for small cell systems," in *Proceedings IEEE 41st Vehicular Technology Conference*, St. Louis, MO, 1991, pp. 589-593.

[12] T. Manabe, Y. Miura, and T. Ihara, "Effects of antenna directivity and polarization on indoor multipath propagation characteristics at 60 GHz," *IEEE Journal on Selected Areas in Communications*, vol. 14, No. 3, pp. 441-447, April 1996.

[13] S. E. Alexander and G. Pugliese, "Cordless communication within buildings: results of measurements at 900 MHz and 60 GHz," *British Telecom Technology Journal*, vol. 44, No. 10, pp. 99-105, October 1996.

[14] Y. P. Zhang, Y. Hwang, "Measurements of the characteristics of indoor penetration loss," *IEEE 44th Vehicular Technology Conference*, vol. 3, pp. 1741-1744, June 1994.

[15] L. Golding and A. Livine, "RLAN-A radio local area network for voice and data communications," *Proceedings IEEE GLOBECOM* vol. 3, pp 1900-1904, November 1987.

[16] T. S. Rappaport, *Wireless Communications: Principles and Practice*, 2nd Edition. New Jersey: Prentice-Hall, 2002.

[17] B. Langen, G. Lober, W. Herzig, "Reflection and transmission behaviour of building materials at 60 GHz," *IEEE 4th International Conference on Personal, Indoor, and Mobile Radio Communications*, vol. 4, pp. 505-509, September 1994.

[18] L. M. Correia and P. O. Francês, "Estimation of materials characteristics from power measurements at 60 GHz," *IEEE 7th International Conference on Personal, Indoor, and Mobile Radio Communications*, vol. 1, pp. 510-513, October 1994.

[19] K. Sato, T. Manabe, T. Ihara, H. Saito, S. Ito, T. Tanaka, K. Sugai, N. Ohmi, Y. Murakami, M. Shibayama, Y. Konishi, and T. Kimura, "Measurements of reflection and transmission characteristics of interior structures of office building in the 60-GHz band," *IEEE Transactions*

*on Antennas and Propagation*, vol. 45, No. 2, pp. 1783-1792, December 1997.

[20] M. Lott and I. Forkel, "A multi-wall-and-floor model for indoor radio propagation," *IEEE 53rd Vehicular Technology Conference*, vol. 1, pp. 464-468, May 2001.

[21] K. Sato, T. Manabe, J. Polivka, T. Ihara, Y. Kasashima, and K. Yamaki, "Measurement of the complex refractive index of concrete at 57.5 GHz," *IEEE Transactions on Antennas and Propagation*, vol. 44, No. 1, pp. 35-40, January 1996.

[22] P. Nobles and F. Halsall, "Indoor propagation at 17 GHz and 60 GHz – measurements and modeling," *IEE National Conference on Antennas and Propagation*, pp. 93 – 96, 1999.

[23] P. Karlsson, C. Bergljung, E. Thomsen, and H. Börjeson, "Wideband measurement and analysis of penetration loss in the 5 GHz band," *Proceedings IEEE 50th Vehicular Technology Conference,* Amsterdam, Netherlands, September 1999, vol. 4, pp. 2323-2328.

[24] C. R. Anderson, "Design and implementation of an ultrabroadband millimeter-wavelength vector sliding correlator channel sounder and in-building measurements at 2.5 & 60 GHz," Masters Thesis, Virginia Polytechnic Institute and State University, http://scholar.lib.vt.edu/theses/index.html, May 2002.

[25] M. J. Feuerstein, K. L. Blackard, T. S. Rappaport, S. Y. Seidel, H. H. Xia, "Path loss, delay spread, and outage models as functions of antenna height for microcellular system design," *IEEE Transactions on Vehicular Technology*, vol. 44, No. 3, pp. 487-497, August 1994.

[26] W. L. Stutzman and G. A. Thiele, *Antenna Theory and Design*, 2nd Edition.  New York: John Wiley & Sons, 1998.

[27] D. M. Matic, H. Harada, R. Prasad, "Indoor and outdoor frequency measurements for MM-waves in the range of 60 GHz," *Proceedings IEEE 48th Vehicular Technology Conference*, vol. 1, pp. 567-571, May 1998.

[28] R. R. Skidmore, T. S. Rappaport, A. L. Abbott, "Interactive coverage region and system design simulation for wireless communication systems in multifloored indoor environments: SMT Plus," *IEEE 5th International Conference on Universal Personal Communications*, vol. 2, pp. 646-650, October 1996.

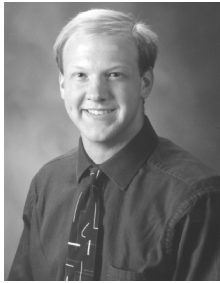

**Christopher R. Anderson (S '96)** received his B.S.E.E. (in Honors) and M.S.E.E from Virginia Tech in 1999 and 2002, respectively.  He is currently working toward the Ph.D. degree at the Mobile and Portable Radio Research Group (MPRG) at the same university as a Bradley Fellow.

Since 1999, he has been a Research Assistant at MPRG, where his research focuses on designing wireless measurement systems as well as studying radiowave propagation.  He has extensive experience performing channel measurements at a variety of frequency bands between 800 MHz and 60 GHz.  In 1996, he participated in a co-op work experience with Corning Cable Company, and has held summer internships with International Resistor Corporation (IRC) in 1999 and Andrew Corporation's Wireless Innovations Group in 2001, 2002, and 2003.

Anderson is a member of the National Society of Professional Engineers and the Radio Club of America.

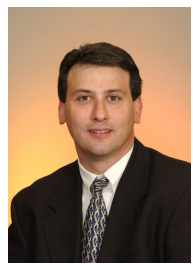

**Theodore S. Rappaport (S'83–M'84–SM'91–F'98)** received the B.S.E.E., M.S.E.E., and Ph.D. degrees from Purdue University, West Lafayette, IN, in 1982, 1984, and 1987, respectively. From 1988 to 2002, he was on the Virginia Tech Electrical and Computer Engineering faculty, where he was the James S. Tucker Professor and founder of the Mobile & Portable Radio Research Group (MPRG), one of the world's first university research and teaching centers dedicated to the wireless communications field. He joined the University of Texas in 2002 as the William and Bettye Nowlin Chair in Engineering, and is director of the newly-formed Wireless Networking and Communications Group (WNCG) at UT's Austin campus. In 1989, he founded TSR Technologies, Inc., a cellular radio/PCS manufacturing firm that he sold in 1993, and founded Wireless Valley Communications, Inc. in 1995. Rappaport received the Marconi Young Scientist Award in 1990, an NSF Presidential Faculty Fellowship in 1992, the Sarnoff Citation from the Radio Club of America in 2000, and the James R. Evans Avant Garde award from the IEEE Vehicular Technology Society in 2002.  He received the Frederick Emmons Terman Outstanding Electrical Engineering Educator Award from the ASEE in November 2002.

Dr. Rappaport has 30 patents issued or pending and has authored, co-authored and co-edited 18 books in the wireless field, including the popular textbooks Wireless Communications: Principles & Practice (Prentice-Hall, 1996, 2002), and Smart Antennas for Wireless Communications: IS-95 and Third Generation CDMA Applications (Prentice Hall, 1999). He has co-authored more than 200 technical journal and conference papers and was recipient of the 1999 IEEE Communications Society Stephen O. Rice Prize Paper Award. Since 1998, he has been series editor for the Prentice Hall Communications Engineering  and Emerging Technologies book series. He serves on the editorial board of International Journal of Wireless Information Networks (Plenum Press, NY) and the advisory board of Wireless Communications and Mobile Computing for Wiley InterScience, is a Fellow of the IEEE, and is active in the IEEE Communications and Vehicular Technology societies. Dr. Rappaport also serves as chairman of Wireless Valley Communications, Inc., an in-building/campus design, measurement, and management company. He is a registered professional engineer in the state of Virginia and is a Fellow and past member of the board of directors of the Radio Club of America. He has consulted for over 25 multinational corporations and has served the International Telecommunications Union as a consultant for emerging nations. He is married and has three children.